\documentclass[preprint,12pt]{elsarticle}

\usepackage{graphicx}
\graphicspath{ {./figures/} }
\usepackage{subfigure}

\usepackage{amsmath,amssymb,amsfonts,amsthm}

\usepackage{textcomp}
\usepackage{xcolor}

\usepackage{hyperref}
\usepackage{mathtools}
\usepackage[Module]{algorithm}
\usepackage{algpseudocode}

\algblockdefx[METHODBLOCK]{Method}{EndMethod}%
	[2]{\textbf{method} \textsc{#1}(#2)}%
	{\textbf{end method}}

\theoremstyle{plain}
\newtheorem{attack}{Attack} 

\newif\ifshowcomments
\showcommentstrue

\ifshowcomments
\newcommand{\mynote}[2]{\fbox{\bfseries\sffamily\scriptsize{#1}}
	{\small$\blacktriangleright$\textsf{\emph{#2}}$\blacktriangleleft$}}
\else
\newcommand{\mynote}[2]{}
\fi

\begin{document}
	
\begin{frontmatter}

\title{Towards a Supply Chain Management System for Counterfeit Mitigation using Blockchain and PUF} 



\author{Leonardo Aniello\corref{cor1}} \ead{l.aniello@soton.ac.uk} 
\author{Basel~Halak} \ead{basel.halak@soton.ac.uk}
\author{Peter~Chai} \ead{xc5g15@ecs.soton.ac.uk}
\author{Riddhi~Dhall} \ead{rd12g15@ecs.soton.ac.uk}
\author{Mircea~Mihalea} \ead{mm8g15@ecs.soton.ac.uk}
\author{Adrian~Wilczynski} \ead{aw11g15@ecs.soton.ac.uk}
\address{Electronics and Computer Science, University of Southampton, United Kingdom}

\cortext[cor1]{Corresponding author}

\begin{abstract}
The complexity of today's supply chain, organised in several tiers and including many companies located in different countries, makes it challenging to assess the history and integrity of procured physical parts, and to make organisations really accountable for their conduct. This enables malicious practices like counterfeiting and insertion of back doors, which are extremely dangerous, especially in supply chains of physical parts for industrial control systems used in critical infrastructures, where a country and human lives can be put at risk.

This paper aims at mitigating these issues by proposing an approach where procured parts are uniquely identified and tracked along the chain, across multiple sites, to detect tampering. 
Our solution is based on consortium blockchain and smart contract technologies, hence it is decentralised, highly available and provides strong guarantees on the integrity of stored data and executed business logic. The unique identification of parts along the chain is implemented by using physically unclonable functions (PUFs) as tamper-resistant IDs.

We first define the threat model of an adversary interested in tampering with physical products along the supply chain, then provide the design of the tracking system that implements the proposed anti-counterfeiting approach. We present a security analysis of the tracking system against the designated threat model and a prototype evaluation to show its technical feasibility and assess its effectiveness in counterfeit mitigation. Finally, we discuss several key practical aspects concerning our solution ad its integration with real supply chains.
\end{abstract}



\end{frontmatter}

\section{Introduction}\label{s:introduction}

Counterfeited products can lead to catastrophic consequences, in particular when such products are used in critical infrastructure, military applications or in food and medicine industries. These include significant  economic losses (e.g. in the order of billion USD per year in the UK~\cite{/content/publication/9789264279063-en}), serious security risks from malfunctioning military weapons and vehicles due to counterfeited parts~\cite{horvath2017not}, and potentially loss  for human lives (e.g. deaths due to contaminated food, such as 2018 E. coli infection~\footnote{Multistate Outbreak of E. coli O157:H7 Infections Linked to Romaine Lettuce (Final Update), available online \url{https://www.cdc.gov/ecoli/2018/o157h7-04-18/index.html}}). It is therefore of paramount importance to develop and deploy effective strategies for counterfeit mitigation to ensure a trustworthy and secure supply chain.  One of main factors magnifying the scale of the counterfeit problem is the trend towards globalisation. The latter is driven by  the need to cut costs to gain a competitive advantage and resulted in a remarkable growth of outsourcing levels, which in turn led to a significant increase of supply chains complexity because more firms are involved and the chain must be spread over further tiers~\cite{wiengarten2016risk}. Such an evolution of the supply chain structure has brought about a number of serious challenges linked to the problem of counterfeiting:

\begin{itemize}
	\item \textbf{Visibility~\cite{hohenstein2015research}.} The network of buyer-supplier relationships has become more intricate and participants have little to no visibility and control on upstream stages, which makes it harder to assess the integrity of procured parts. 
	\item \textbf{Traceability~\cite{khojasteh2018supply}.} Tracking data is fragmented and spread among involved companies, which makes it very challenging to uniquely identify each procured item and trace its history back to its origin and, in case of incidents, there is a shortage of data that can be used for forensics investigations.
	\item \textbf{Accountability~\cite{HARTMANN2014281}.} In such a scenario afflicted by obscurity and lack of information, fraudulent conduct of companies is noticeably facilitated. There is a lack of means to keep organisations accountable for the portion of processing they handle within the supply chain. 
\end{itemize}

Coping with counterfeiting in these supply chains calls for a platform integrated throughout the whole chain to reliably record every transition of products between involved companies. The availability of such a ledger would be an effective means to provide any legitimate actor with precise information on what organisations are operating at upstream stages of the chain (\textit{visibility}) and on the history of each procured item (\textit{traceability}). Moreover, ensuring recorded transactions are truthful and not tampered with is crucial to enable legally binding liability policies (\textit{accountability}). The implementation of such a platform for counterfeit mitigation requires an infrastructure deployed over the considered supply chain, to enable fine-grained monitoring of physical parts sold and bought by involved companies. It would be infeasible to identify a single specific authority or enterprise eligible for controlling and operating an infrastructure like this, possibly spanning different countries and diverse regulatory frameworks. Furthermore, such an authority should be trusted globally and have the resources to effectively setup and maintain such a world-wide, complex interconnected network, ensuring at the same time top levels of security, availability and performance. 

A decentralised approach is more suitable, where the infrastructure itself is a peer-to-peer network distributed across all the supply chain partners, devoid of any centralised control that may become a single point of failure or a performance bottleneck. An emerging technology that lends itself well to implement a platform like that is the \textit{blockchain}, because of its full decentralisation, high availability and strong guarantees on the immutability of stored data. In brief, a blockchain is a distributed system consisting of a network of peer nodes sharing a ledger of transactions, where each peer keeps a replica of that ledger. The consistency among replicas is ensured by a distributed consensus algorithm run by all the nodes, which also guarantees that transactions cannot be censored or redacted unless an attacker succeeded in controlling a certain percentage of nodes or of computational power. In addition to storing data, blockchain can be used to execute application logic through the \textit{smart contract} technology. A smart contract is an application whose code and execution traces are stored immutably in the blockchain, which provides strong guarantees on execution integrity.

Since such infrastructure has to be run across a predefined set of parties, and considering that part of managed data is not meant to be disclosed publicly, it is reasonable to not rely on existing public permissionless blockchains like Ethereum's. Rather, it is more sensible to build on a \textit{consortium blockchain} where nodes are authenticated, membership is predetermined and data cannot be accessed from the outside. 

In this paper, \textit{we propose an approach for item tracking in supply chains based on consortium blockchain and smart contract technologies}. Items are uniquely identified to enable tracking by using tamper-proof tags. We choose to use \textit{physically unclonable functions} (PUF) to implement those tags. PUFs are circuits that provide unique signatures deriving from manufacturing process variations of the circuits themselves. Each alteration of those tags leads to changes of the function computed by the PUF, hence this technology is well suited to enable counterfeit detection. We provide the design of a supply chain management system based on the proposed approach and carry out a preliminary analysis on its effectiveness and feasibility. We define the adversary model to characterise what types of threats can arise in the context of supply chain counterfeit. We then analyse how the proposed design can address those threats to deliver improved counterfeit detection. Finally, to show the technical feasibility of this solution, we describe its prototype implementation and preliminary experimental evaluation, where we measure the effectiveness of using PUFs for counterfeit detection. Finally, we provide an ample discussion on some key pragmatic aspects of integrating the proposed platform with real supply chains. 

Although some other blockchain-based supply chain management systems have been proposed in literature and industry, a few of them rely on PUFs for item tracking. The main novelty of this work lies in the level of detail of the proposed design, the corresponding security analysis and prototype implementation and evaluation.

\medskip

\noindent \textbf{Our Contribution.}
In this paper, we rely on blockchain, smart contract and PUF technologies to design a tracking system of physical parts for supply chain management, aimed at mitigating the problem of counterfeiting. With respect to the state of the art on this topic, our main research contributions are
\begin{itemize}
	\item the explicit modelling of the overall system, including supply chain, blockchain, smart contracts, PUFs and adversary behaviour, i.e. the \textit{threat model};
	\item the detailed design of the proposed tracking system for detecting counterfeits in supply chains;
	\item based on the designated threat model, the identification of the possible attacks to the tracking system aimed to bypass counterfeit detection;
	\item the analysis of how the proposed tracking system reacts against each of the identified attacks;
	\item a prototype implementation and preliminary experimental evaluation of the proposed tracking system, where PUF-based counterfeit detection accuracy is assessed;
	\item a discussion on most relevant points concerning the integration of our solution in real scenarios.
\end{itemize}

\medskip

\noindent \textbf{Paper Organisation.}
The remainder of this paper is organised as follows. Section~\ref{s:related_work} describes related work. Section~\ref{s:preliminaries} introduce background information on PUF, blockchain and smart contract technologies. The system model is presented in section~\ref{s:system_model}, as well as the threat model. Our tracking system is detailed in section~\ref{s:tracking} and its security properties are analysed in section~\ref{s:security_analysis}. Section~\ref{s:evaluation} describes the prototype implementation and evaluation. Section~\ref{s:discussion} discusses security analysis results and the limitations of our solution. Finally, section~\ref{s:conclusion} outlines conclusion and future work.

\section{Related Work} \label{s:related_work}

The use of blockchain and smart contracts for supply chain management is currently being investigated in some recent industrial projects~\footnote{How Blockchain Will Transform The Supply Chain And Logistics Industry (\url{https://www.forbes.com/sites/bernardmarr/2018/03/23/how-blockchain-will-transform-the-supply-chain-and-logistics-industry})}~\footnote{Using blockchain to drive supply chain transparency (\url{https://www2.deloitte.com/us/en/pages/operations/articles/blockchain-supply-chain-innovation.html})}, and led to the launch of a number of new businesses and companies, which supports the perceived potentialities of this application. Some of these projects use a blockchain-as-a-service solution provided by a third party, such as TradeLends~\footnote{TradeLends, available online \url{https://www.tradelens.com/}}, which employs the platform delivered by IBM Cloud. The limitation of such an approach is the need to totally trust an external organisation, which brings about the same issues mentioned before regarding centralisation.

Different companies use diverse technologies to tag products and reliably link physical assets to the blockchain. WaBi~\cite{wabi} and Waltonchain~\cite{waltonchain} use RFID (Radio-frequency identification) as tags to identify and track items along the chain. Others make use of proprietary solutions. For example, BlockVerify~\cite{blockverify} uses their own Block Verify tags, Chronicled~\cite{chronicled} employs trusted IoT chips, Skuchain~\cite{skuchain} applies Proof of Provenance codes called Popcodes. The problem of existing approaches that rely on the use of RFID-based tags is that these tags are vulnerable to cloning attacks~\cite{jain2018analysis,7005493}, this makes it less effective in protecting against counterfeit attempts.

\textit{We propose to produce tamper-proof tags by using physically unclonable functions (PUF)}, 
i.e. circuits that can generate a unique identifier for each chip due to the intrinsic variability of the IC fabrication process. 
Previously reported works on using PUF technology in the context of supply chain management are limited in both scope and depth. Guardtime~\cite{guardtime} proposes the use of PUF for IoT device authentication, but provides no clear information on the integration with supply chain. Islam et al.~\cite{islam2018ic} propose the use of PUF for tracing integrated circuits (IC). Their work is focussed on hardware-supply chain and does not investigate in depth what security guarantees are provided. To the best of our knowledge, the latter aspect, i.e. the lack of appropriate security analysis of the proposed solution, is currently a gap in the state of the art on the application of blockchain and smart contract technologies for counterfeiting mitigation in supply chains.

Alzahrani and Bulusu~\cite{Alzahrani:2018:BCN:3211933.3211939} present Block-Supply Chain, a design for a blockchain-based supply chain where products are tracked using Near Field Communication (NFC) technology to detect counterfeits. Their security analysis is limited to the novel consensus protocol they propose and does not take into account any other aspect of the overall supply chain ecosystem, which includes, but is not restricted to, the blockchain. Furthermore, they do not define a threat model to specify what attacks they want to defend from.

Toyoda et al.~\cite{7961146} introduce a blockchain-based solution for product ownership management system, to be used to prevent counterfeits in the post supply chain. They explain how their system allows to detect counterfeits, and discuss the provided security guarantees only in terms of the possible vulnerabilities of the underlying technology they use, i.e. Ethereum~\footnote{Ethereum Project (\url{https://www.ethereum.org/})}. Also in this case, there is no explicit modelling of the adversary to identify possible attacks and analyse how the proposed solution copes with them.

Similarly, Negka et al.~\cite{negka2019employing} describes a method to detect counterfeit IoT devices by tracking each single device component along the supply chain. They rely on PUFs to authenticate components and implement their detection logic in Ethereum. Although they provide some figures on the fees to pay to use Ethereum smart contracts, they do not detail how PUFs and smart contracts are integrated, nor what specific mechanism is actually employed to implement the detection. Obtained detection accuracy and provided security guarantees are not discussed.

\section{Preliminaries} \label{s:preliminaries}

In this section we introduce some preliminary background on physically unclonable functions (section~\ref{s:puf}) and blockchain and smart contract technology (section~\ref{s:blockchain}).



\subsection{Physically Unclonable Function} \label{s:puf}

Physically unclonable functions (PUF) are security primitive capable of generating a hardware-based digital signature unique for each device~\cite{halak2016overview}. 
PUFs are commonly implemented as circuits and ensure that responses are each hardware by exploiting the inherent randomness of the internal structure introduced by the manufacturing process.
This technology has many attractive advantages, including its relatively low cost (a typical PUF can be built using few thousands transistors), and its inherent security deriving from the extreme difficulty of forging its design. Indeed, it is almost impossible to create a physical clone of a PUF, which means that this technology can be used reliably to identify those physical objects where a PUF can be integrated, and therefore to detect possible forgery.  
From a mathematical point of view, a PUF is a function that generates an output (also called \textit{response}) starting from an input (also called \textit{challenge}). The challenge-response data (CRD) must be unique for a single device. 
The use of PUF for building entity-authentication protocols has been extensively explored in the literature~\cite{yilmaz2018lightweight, chatterjee2018building, wild2017fair}. In general, each entity is provided with a PUF and the authentication scheme consists of two stages~\cite{halak2018hardware}: 
\begin{enumerate}
	\item \textit{Enrolment Phase:} when a new entity has to be enrolled, a verifier collects the required CRD from entity's PUF and stores it in a database, together with the ID of the entity itself.
	\item \textit{Verification Phase:} when an enrolled entity has to be authenticated, the verifier receives the entity ID and retrieves the corresponding CRD from the database. A random challenge-response pair is selected form the CRD and the challenge is sent in clear to the entity, which computes the response by using its PUF and sends it back in clear to the verifier. If the response corresponds to that stored in the database, then the authentication is successful and the challenge-response pair is removed from stored CRD to prevent replay attacks. Otherwise, the authentication fails. 
\end{enumerate}

Ideally, a PUF should always generate the same response for a given challenge. Unfortunately, conditions such as temperature or voltage variations could lead to different responses~\cite{halak2018security}.

A PUF can be implemented in different ways and with different technologies, leading to varying security guarantees. For example, PUFs based on SRAM have been proved to be clonable~\cite{halak2018security}, which questions their suitability to be used to implement authentication protocols. It has been also shown that a PUF can be vulnerable to machine learning (ML)-based modelling attacks~\cite{halak2018security}, where an adversary builds an accurate mathematical model of the PUF by collecting a sufficient number of challenge-response pairs, and uses that model to clone the PUF itself. There are a number of techniques that can be used to mitigate the risks of ML-based attacks, such as using cryptographic blocks to obfuscate the output of the PUF~\cite{mispan2018cost}, increasing the circuit complexity of the design~\cite{su2018machine}, or solving this issue at the protocol level~\cite{yilmaz2018lightweight, yu2016lockdown}. 


\subsection{Blockchain and Smart Contract} \label{s:blockchain}


A blockchain is a ledger of transactions, replicated among a number of nodes organised in a peer-to-peer network. Transactions are submitted to the blockchain network and stored in the ledger. A consensus algorithm is run among blockchain nodes to guarantee the consistency of the ledger, in terms of what transactions are included in which order. A blockchain provides strong guarantees in terms of availability, because a peer-to-peer network with several nodes and no single-point-of-failure is used. Furthermore, as the ledger is replicated and several nodes participate in the consensus algorithm, an adversary should take control of a relevant fraction of nodes to take over the blockchain and tamper with the ledger. That fraction of nodes depends on the chosen consensus algorithm.

In open, \textit{permissionless blockchains} like Bitcoin’s~\footnote{Bitcoin (\url{https://bitcoin.org/en/})} and Ethereum’s, any node can join the network without any form of authentication, hence additional mechanisms are required to cope with the potential presence of malicious nodes. Proof-of-Work (PoW) is commonly employed, which, although effective in countering cyber threats stemming from malicious blockchain nodes, is time-consuming and greatly restricts performance~\cite{DBLP:conf/itasec/GaetaniABLMS17}. In \textit{consortium blockchains} like Hyperledger Fabric~\footnote{Hyperledger Fabric (\url{https://www.hyperledger.org/projects/fabric})}, blockchain membership is restricted to the nodes owned by interested organisations, so that each involved firm can take part to the overall process and no external actor can interfere with any operation or read any exchanged data. In this way, blockchain nodes are known and can be reliably authenticated, which allows to replace PoW with other, more efficient techniques that ensure high level performance in terms of latency and throughput, such as byzantine fault tolerance algorithms~\cite{Castro:1999:PBF:296806.296824}.

On top of a blockchain, a smart contract execution environment can be built, to extend the functionalities of the blockchain beyond storing data and allow the execution of any application logic. A smart contract is the code implementing the required application logic and it can be installed in a bockchain likewise a normal transaction, which ensures consequently its integrity. A smart contract defines an interface with methods that can be called externally. Each invocation of a smart contract method is stored as a blockchain transaction, hence the execution trace can be considered as immutable. In general, computations executed through smart contracts are fully transparent and tamper-proof.


\section{System Model} \label{s:system_model}

This section defines the system model representing supply chain (section~\ref{s:supply_chain_model}), PUF-equipped items (section~\ref{s:puf_model}), blockchain and smart contracts (sections~\ref{s:blockchain_model} and~\ref{s:smart_contract_model}, respectively). Finally, thread model is introduced in section~\ref{s:threat_model}.

\subsection{Supply Chain Model} \label{s:supply_chain_model}
A supply chain $\mathcal{SC}$ includes $N$ parties $\mathcal{P} = \{p_i\}$, i.e. organisations involved in the chain with different roles, and that engage among themselves by supplying and buying items. A \textit{supplier} is a party that provides items, while a \textit{buyer} is a party that receives items. Each party can act at the same time as supplier for a number of buyers and as buyer for diverse suppliers. There can be parties that are neither suppliers nor buyers for any other party but operate anyway in the supply chain, such as auditors or regulators. This kind of parties usually needs to access tracking data to assess compliance and solve disputes.

\begin{figure}[h]
	\includegraphics[width=8cm]{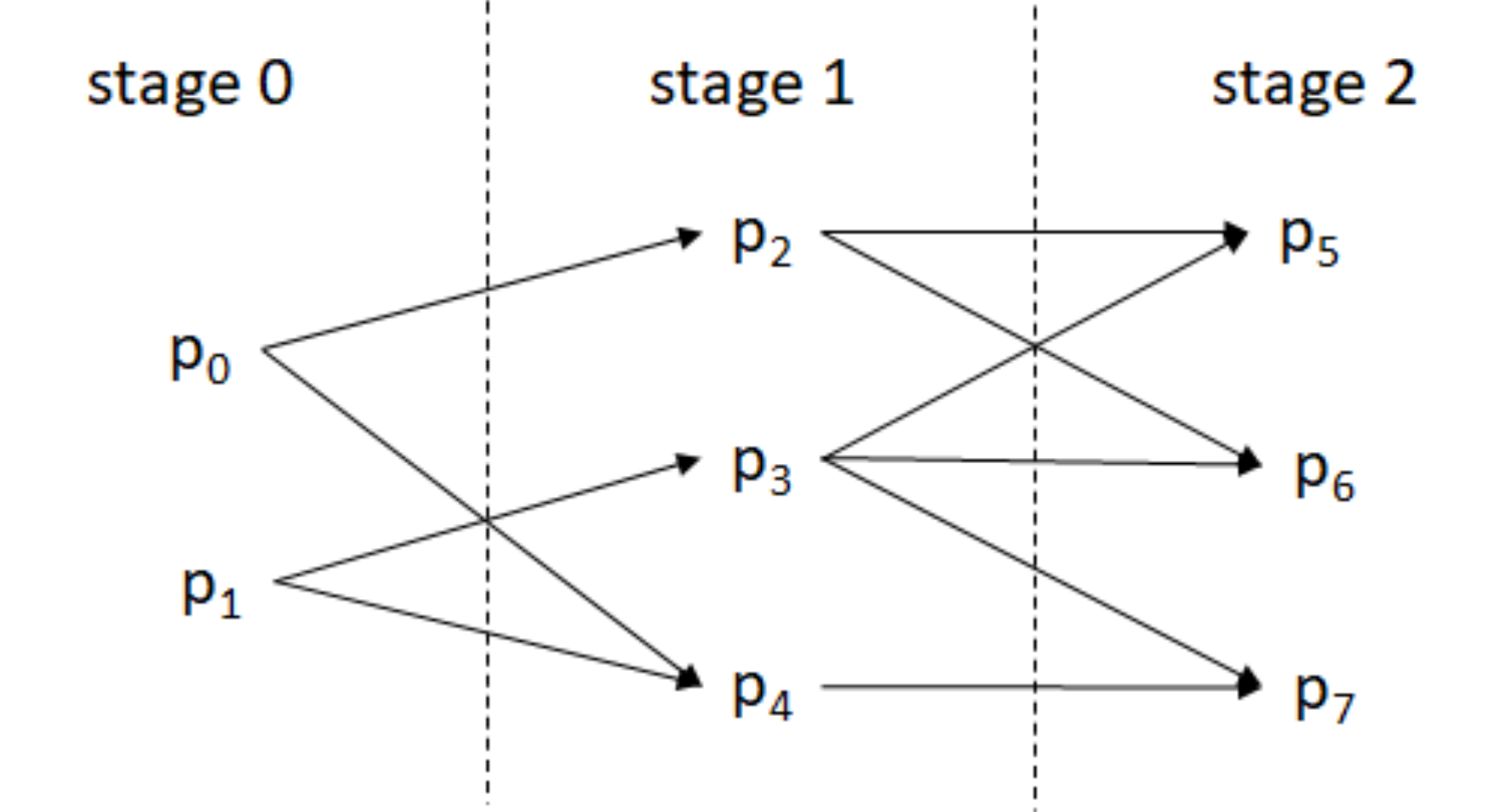}
	\centering
	\caption{Graphical representation of an instance of the supply chain model with 8 parties $p_0, \dots, p_7$ spread across 3 stages. Arrows represent the supplier-buyer relationships, e.g. $(p_1, p_3)$ models the fact that $p_1$ is a supplier of $p_3$.}
	\label{f:supply_chain_model_instance}
\end{figure}

We model $\mathcal{SC}$ as a directed acyclic graph $(\mathcal{P},\mathcal{R})$, where $\mathcal{R}$ is the set of binary supplier-buyer relationships holding within $\mathcal{SC}$. Figure~\ref{f:supply_chain_model_instance} shows an instance of the supply chain model. Each element of $\mathcal{R}$ is in the form $(p_i, p_j)$, with $p_i,p_j \in \mathcal{P} \land p_i \ne p_j$, and represents a supplier-buyer relationship where $p_i$ is the supplier and $p_j$ the buyer. According to these relationships, parties can be organised in stages, i.e. the stages of the supply chain. Let $S$ be the number of stages of $\mathcal{SC}$. Without loss of generality~\footnote{It would be possible for an organisation to operate at different stages of a supply chain. In these cases, we model such an organisation as multiple parties, one for each stage where it operates.}, we define the function $stage\colon \mathcal{P} \to \mathbb{N}$ as follows
\begin{equation}
	stage(p) =
	\begin{cases*}
		0 & iff $\nexists q \in \mathcal{P} \mid (q,p) \in \mathcal{R}$ \\
		i+1 & otherwise, where $i = \max\limits_{q \in \mathcal{P} \mid (q,p) \in \mathcal{R}} stage(q)$
	\end{cases*}
	\label{eq:stage}
\end{equation}

Although equation~\ref{eq:stage} covers the cases where a buyer has suppliers in different stages, this is not likely to happen in real supply chains. 
Indeed, buyers commonly purchase items from parties in the previous stage only. Therefore we introduce the following constraint
\begin{equation}
	\forall (p,q) \in \mathcal{R} \quad stage(q) - stage(p) = 1
\end{equation}

We assume the existence of a reliable public key infrastructure (PKI) for the parties in $\mathcal{P}$. Each party $p_i$ has a key pair $(pk_i, sk_i)$, where $pk_i$ is the public key known to all the other parties and $sk_i$ is the private key known to $p_i$ only. We discuss in section~\ref{s:discussion} how such a PKI can be realised and the related issues. Given a key $k$ and a plaintext message $m$, we indicate with $\lvert m \rvert_k$ the ciphertext derived from encrypting $m$ with $k$. We use $\langle m \rangle_{\sigma_i}$ to indicate that the message $m$ has been signed by $p_i$, i.e. that it includes a digest of $m$ encrypted with $sk_i$.

\subsection{PUF-equipped Item Model} \label{s:puf_model}

\begin{figure}[h]
	\includegraphics[width=8cm]{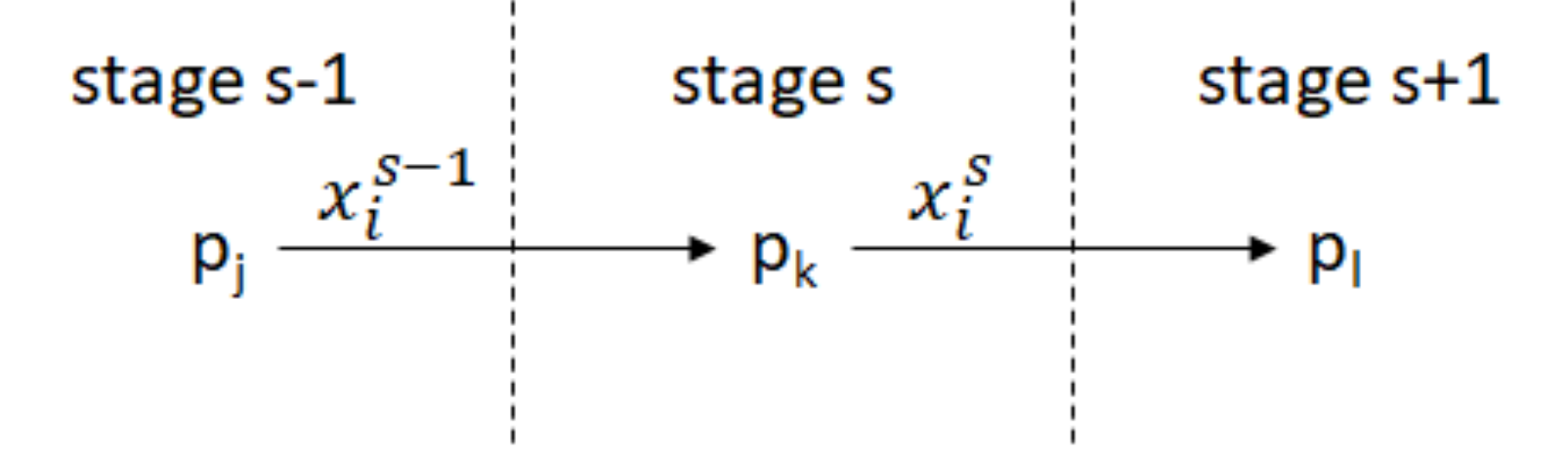}
	\centering
	\caption{Graphical representation of item $x_i$ moving from $p_j$ in stage $s-1$ to $p_k$ in stage $s$ ($x_i^{s-1}$), and from there to $p_l$ in stage $s+1$ ($x_i^{s}$).}
	\label{f:item_moving}
\end{figure}

A number of items are moved along the supply chain $\mathcal{SC}$, from parties at stage 0 to downstream parties. We refer to the generic $i^{\text{th}}$ PUF-equipped item produced at stage $0$ of the supply chain as $x_i$. Furthermore, as items can be forged along the chain, we define $x_i^s$ as the item $x_i$ \textit{after its processing} at stage $s$, where $s = 0 \dots S-1$. That is, $x_i^s$ is the item $x_i$ when it is delivered from the supplier at stage $s$ to the buyer at stage $s+1$ (see figure~\ref{f:item_moving}).

We refer to the function computed by the PUF integrated with item $x_i^s$ as $puf_i^s\colon \mathbb{N} \to \mathbb{N}$. 
When an item $x_i$ is produced at stage $0$ and equipped with a PUF, it is considered intact.
If $x_i$ is never tampered with along the chain, then the following property holds with high probability~\footnote{as explained in section~\ref{s:puf}, the function computed by a PUF is not 100\% stable}
\begin{equation}
	\label{eq:tampering_check_positive}
	\forall c \in \mathbb{N} \quad \forall s \in [1 \dots S-1]  \quad puf_i^0(c) = puf_i^s(c)
\end{equation}
If instead $x_i$ is forged at stage $s > 0$, then $puf_i^0 \ne puf_i^s$ and the following property holds with high probability~\footnote{even if the two functions are different, they might return the same response for some challenge}
\begin{equation}
	\label{eq:tampering_check_negative}
	\forall c \in \mathbb{N} \quad puf_i^0(c) \ne puf_i^s(c)
\end{equation}

We consider the case where PUFs are built by using techniques that mitigate the risk of ML-based attacks~\ref{s:puf}, hence we assume that an adversary cannot clone a PUF by collecting a sufficient number of challenge-response pairs. 


\subsection{Blockchain Model} \label{s:blockchain_model}
We consider a \textit{consortium blockchain} $\mathcal{B}$ with $N$ nodes $\mathcal{N} = \{n_i\}$, deployed over the supply chain parties' premises (see section~\ref{s:blockchain}). More precisely, node $n_i$ is located at party $p_i$. Nodes can communicate among each other over the network by sending messages. The network is asynchronous, there is no known bound on message latencies but messages are eventually delivered to their destination. $\mathcal{B}$ uses a \textit{byzantine fault tolerant consensus} protocol, such as PBFT~\cite{Castro:1999:PBF:296806.296824}, which ensures \textit{safety} if up to $f = \lfloor \frac{N-1}{3} \rfloor$ nodes are byzantine. Subsection~\ref{s:threat_model} will explain how byzantine nodes behave. 

Interactions between nodes take place by sending digitally signed messages. When a node $n_i$ wants to send a message $m$ to another node $n_j$, $n_i$ sends a message $\langle i, j, ts, m \rangle_{\sigma_i}$ to $n_j$. The parameter $ts$ is a timestamp set by $n_i$, used to avoid replay attacks.

Clients running within supply chain parties' premises can submit transactions to $\mathcal{B}$ by broadcasting them to all $\mathcal{B}$'s nodes. Submitted transactions are eventually confirmed by $\mathcal{B}$ and persistently stored, with strong guarantees on their immutability, i.e. persisted transactions cannot be tampered with or removed unless more than $f = \lfloor \frac{N-1}{3} \rfloor$ nodes are byzantine.


\subsection{Smart Contract Execution Environment Model} \label{s:smart_contract_model}
Consortium blockchains like those described in section~\ref{s:blockchain_model} can support the execution of smart contracts (see section~\ref{s:blockchain}), i.e. a smart contract execution environment $\mathcal{SCEE}$ can be built on top of a consortium blockchain $\mathcal{B}$. $\mathcal{SCEE}$ is deployed over the same nodes $\mathcal{N}$ of $\mathcal{B}$.

Smart contracts can be installed in $\mathcal{SCEE}$. A smart contract $\mathcal{C}$ includes a number of methods, which can be invoked externally, and a key-value store $kvs$, which can be accessed internally only, inside those methods. The installation of a smart contract $\mathcal{C}$ in $\mathcal{SCEE}$ and every invocation of $\mathcal{C}$'s methods are persisted as transactions submitted to the underlying blockchain $\mathcal{B}$. This implies that the application logic encoded by a smart contract cannot be tampered with as long as the underlying blockchain $\mathcal{B}$ guarantees immutability, i.e. unless more than $f = \lfloor \frac{N-1}{3} \rfloor$ nodes are byzantine.

The key-value store of each smart contract provides an interface $set(k,v)$ and $get(k)$ to set and get values for given keys, respectively. Any internal key-value storage $kvs$ relies on the underlying blockchain $\mathcal{B}$ to ensure consistency and immutability of its state. In the specific, each set operation invoked through the $set(k,v)$ method is saved as a transaction in $\mathcal{B}$, hence the whole redo log of the storage is persisted immutably~\cite{DBLP:conf/itasec/GaetaniABLMS17}. Furthermore, we assume that a single set operation is allowed for each key, i.e. the value stored for a key cannot be overwritten. In case of overwriting attempt, the set operation returns an error.

\subsection{Threat Model} \label{s:threat_model}
The final goal of the adversary is to tamper with items for its own benefit. Hence, it aims at avoiding that counterfeit items are detected to prevent raising suspicion, and anyway it seeks to impede that any detected forgery is attributed to itself. We assume the existence of a single adversary in the supply chain, section~\ref{s:discussion} encompasses a brief discussion on considering the presence of more independent adversaries.

At \textit{supply chain level} (see subsection~\ref{s:supply_chain_model}), the adversary can operate at one of the parties, say $p_A$ at stage $stage(p_A)$, with $A \in [0 \dots N-1]$. We assume that the adversary cannot control more than one party and cannot alter any supplier-buyer relationship. 

At \textit{item level} (see subsection~\ref{s:puf_model}), the adversary can tamper with items during the manufacturing processes of the party $p_A$ where it operates. For each bought item $x_i^{stage(p_A)-1}$, 
the adversary can decide whether or not to forge it before supplying it in turn to some other party. 
However, any tampering with $x_i^{stage(p_A)-1}$ affects the internal structure of the integrated PUF, hence $puf_i^0 \ne puf_i^{s_A}$ (see equations~\ref{eq:tampering_check_positive} and~\ref{eq:tampering_check_negative}). Furthermore, if the adversary succeeds to collect at least $N_{PUF}$ challenge-response pairs, it can build a clone of the PUF and attach it to a different item, i.e. it can replace an original product with a counterfeit. 

At \textit{blockchain and smart contract execution environment levels} (see subsections~\ref{s:blockchain_model} and~\ref{s:smart_contract_model}), the adversary can control the local node $n_A$ of $\mathcal{B}$ and $\mathcal{SCEE}$, i.e. such node is byzantine. The behaviour of a byzantine node can deviate arbitrarily from the expected conduct, hence it can for example drop messages and send not expected or wrong messages. Anyway, the adversary cannot break used cryptographic protocols, hence it cannot decrypt messages encrypted without knowing the corresponding keys and cannot forge message signatures.


\section{Tracking System} \label{s:tracking}

Items are tracked as they move along the supply chain, first when they are produced at stage $0$ and then each time they are supplied to a buyer operating at the next stage. When delivered at buyer side, the integrity of each item is verified by using its integrated PUF. Tracking information are stored as blockchain transactions to ensure they are immutable and available to any party in $\mathcal{P}$.

\begin{figure*}[h]
	\includegraphics[width=12cm]{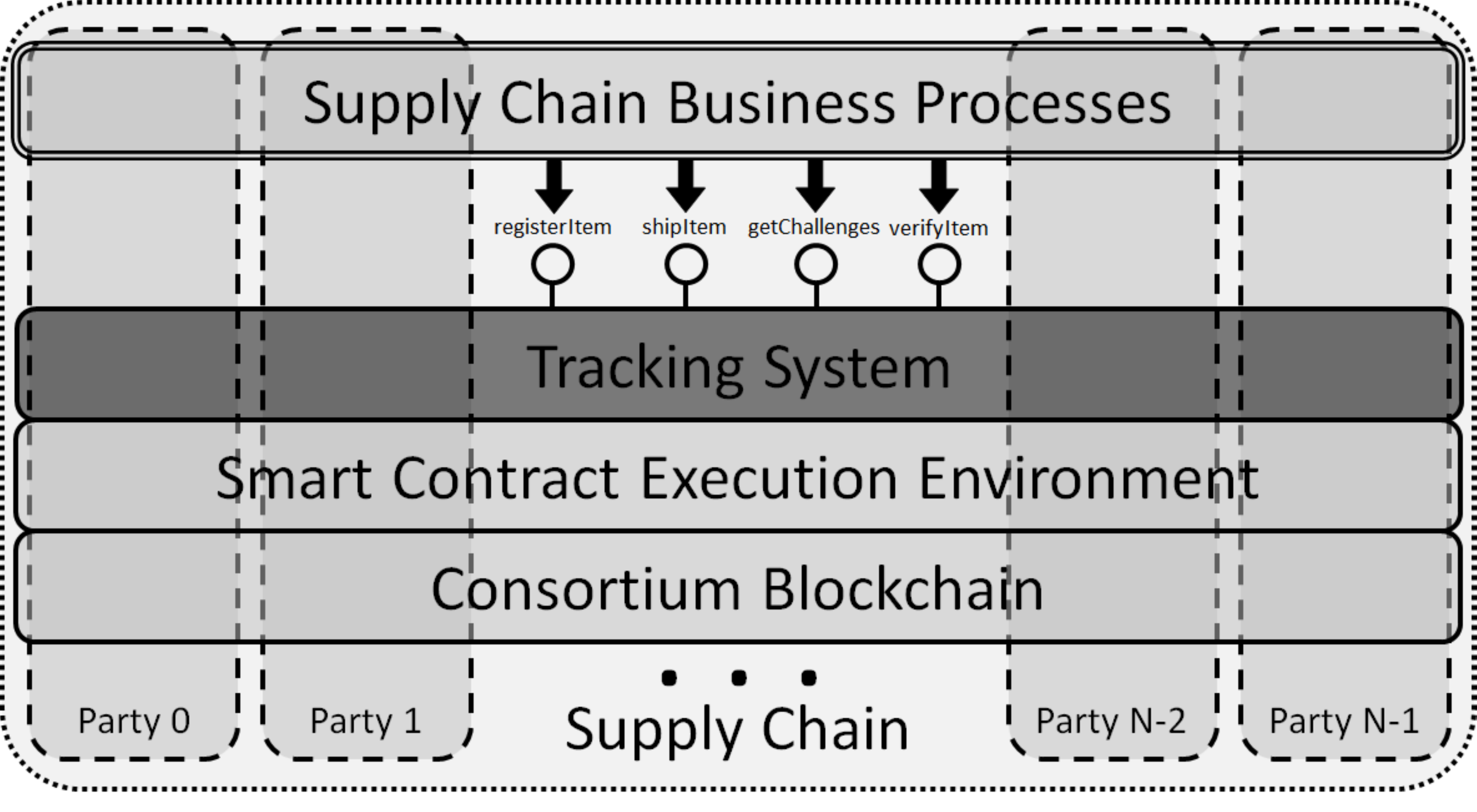}
	\centering
	\caption{High-level architecture of the tracking system and its positioning within the supply chain.}
	\label{f:architecture}
\end{figure*}

The tracking system is built as a smart contract $\mathcal{TS}$ on top of a blockchain-based smart contract execution environment $\mathcal{SCEE}$ (see subsection~\ref{s:smart_contract_model}). 
We consider a consortium blockchain $\mathcal{B}$ like the one presented in subsection~\ref{s:blockchain_model}, and leverage on the PUFs integrated with the items to assess whether they have been tampered with (see subsection~\ref{s:puf_model}). The high-level architecture is shown in figure~\ref{f:architecture}, where basic building blocks and interfaces with supply chain business processes are highlighted. Consortium blockchain $\mathcal{B}$, smart contract execution environment $\mathcal{SCEE}$ and tracking system $\mathcal{TS}$ are distributed and deployed over the IT infrastructures of all the parties. 

\begin{algorithm*}
	\scriptsize
	\caption{Tracking Mechanism} \label{m:tracking_system}
	\begin{algorithmic}[1]
		\Statex \textbf{global variables}: 
		\State $C$ \Comment{number of challenges to send for each verification}
		\State $R$ \Comment{number of responses (out of $C$) that need to be correct for the verification to succeed}
		\Statex
		
		\Statex \textbf{local variables}: 
		\State $kvs$ \Comment{local key-value storage instance}
		\State $p$ \Comment{party's identifier, i.e. this instance is deployed on party $p_p$'s premises}
		\Statex

		\Statex $\vartriangleright$ \textit{$i$ is the item's identifier, $crd_i$ its challenge-response data}
		\Method{registerItem}{$i, crd_i$} 
		\State $kvs.set(\langle crd, i  \rangle_{\sigma_p}, \langle p, crd_i \rangle_{\sigma_p})$ \label{l:crd} \label{l:set_crd}
		\EndMethod
		\Statex

		\Statex $\vartriangleright$ \textit{$b$ is the buyer's identifier, $i$ the item's identifier}
		\Method{shipItem}{$b, i$}
		\State $kvs.set(\langle shipped, p, b, i  \rangle_{\sigma_p}, i)$ \label{l:ship}
		\EndMethod
		\Statex

		\Statex $\vartriangleright$ \textit{$s$ is the supplier's identifier, $i$ the item's identifier}
		\Method{getChallenges}{$s, i$}
			\If{$kvs.get(\langle shipped, s, p, i  \rangle) = null$} \label{l:check_ship} \Comment{check first if the shipping has been tracked}
				\State $kvs.set(\langle no\_ship, s, p, i  \rangle_{\sigma_p}, i)$ \label{l:no_ship}
			\Else
				\State $crd_i \gets \pi_2(kvs.get(\langle crd, i  \rangle))$ \label{l:get_crd}
				\If{$cdr_i = null$} \label{l:check_crd} \Comment{check if the challenge-response data has been stored}
					\State $kvs.set(\langle no\_crd, s, p, i  \rangle_{\sigma_p}, i)$ \label{l:no_crd}
				\Else
					\State $crv \gets \overline{crd_i[p]}$ \Comment{retrieve from $kvs$ and decrypt the $C$ challenge-response pairs} \label{l:select_challenge}
					\State $kvs.set(\langle declare\_verification, s, p, i \rangle_{\sigma_p}, \langle crv \rangle_{\sigma_p})$ \label{l:declare_verification}
				\EndIf
			\EndIf
		\EndMethod
		\Statex

		\Statex $\vartriangleright$ \textit{$s$ is the supplier's identifier, $i$ the item's identifier, $crv$ are the $C$ challenge-response pairs $\langle c_k, r_k \rangle$ previously extracted from $crd_i$, and $crv'$ are the $C$ challenge-response pairs $\langle c_k, r_k' \rangle$ computed by the PUF}
		\Method{verifyItem}{$s, i, crv, crv'$}
			\State $r \gets |\{c_k \mid r_k = r_k'\}|$ \Comment{count how many responses match}
			\If{$r \ge R$}
				\State $kvs.set(\langle verification\_succeeded, s, p, i \rangle_{\sigma_p}, \langle crv \rangle_{\sigma_p})$ \label{l:succeeded}
			\Else
				\State $kvs.set(\langle verification\_failed, s, p, i \rangle_{\sigma_p}, \langle crv, crv' \rangle_{\sigma_p})$ \label{l:failed}
			\EndIf
		\EndMethod
	\end{algorithmic}
\end{algorithm*}

Module~\ref{m:tracking_system} shows the pseudo-code of the tracking system, which defines the four methods shown in figure~\ref{f:architecture}. These methods are used to integrate the proposed tracking mechanism with the business processes of the supply chain. 
In particular, this integration occurs 
on three specific events: when an item is first introduced in the supply chain at stage $0$ (\textit{event 1}, see subsection~\ref{s:event1}), when a supplier ships an item to a buyer (\textit{event 2}, see subsection~\ref{s:event2}) and when a shipped item is delivered to a buyer (\textit{event 3}, see subsection~\ref{s:event3}). After an item has been processed by a party in the last stage, no further tracking is enforced.

All tracking data is kept in the blockchain-based key-value storage via set operations, where any relevant information is digitally signed (see section~\ref{s:supply_chain_model}) by the party executing the method where the set operation itself is invoked. This, together with the constraint that keys cannot be overwritten (see section~\ref{s:smart_contract_model}), ensures that an adversary cannot execute any tracking system method on behalf of another party.

It is to note that, to integrate in practice supply chain $\mathcal{SC}$ business processes with the tracking system $\mathcal{TS}$, an additional layer is required to interface the existing legacy business process management software of $\mathcal{SC}$ with the $\mathcal{TS}$ smart contract. This integration can be achieved through standard software engineering approaches and does not entail any element of novelty or challenge, so it is not described here because out of the scope. Anyway, such integration layer needs to be accounted for as another potential attack surface that the adversary may exploit, hence in section~\ref{s:security_analysis} we also address the corresponding security implications (attack~\ref{a:methods}).


\subsection{Event 1: New Item} \label{s:event1}
When a new item $x_i$ is produced by a party $p_j$ at stage $0$, a PUF is integrated with $x_i$ and a number $C \cdot N$ of challenge-response pairs $\langle c_k , r_k \rangle$ is collected. 
This amounts to $C$ pairs for each party, which allows to use more challenges to verify the integrity of an item at delivery time. In the specific, in order to make the verification process more robust against possible variations in the responses generated by a PUF (see section~\ref{s:puf}), $C$ challenges are used by a buyer when a new item is delivered. 
Each pair is produced by generating a unique random challenge $c_k \in \mathbb{N}$, giving it as input to the PUF of $x_i$ and recording the corresponding output $r_k = puf_i^0(c_k)$. 
The set of pairs is partitioned in $N$ disjoint subsets $ps_w$, with $w = 0 ... N-1$, each containing $C$ pairs.
Every pairs subset is then encrypted with the public key of a different party. 
We refer to the vector of encrypted pairs subsets as the \textit{challenge-response data (CRD)} $crd_i$ of $x_i$, i.e. $crd_i = [ \lvert ps_0 \rvert_{pk_0}, ..., \lvert ps_{N-1} \rvert_{pk_{N-1}}]$. 
The encrypted challenge-response pairs subset $\lvert ps_w \rvert_{pk_w}$ for party $p_w$ is indicated as $crd_i[w]$, and its decrypted version, available for party $p_w$ only, is referred to as $\overline{crd_i[w]}$.

After the generation of the CRD, the method $registerItem(i, crd_i)$ is invoked. This method simply stores in the key-value storage the information that the CRD $crd_i$ for item $x_i$ is available and has been produced by party $p_j$ (line~\ref{l:crd}).

Party $p_j$ has to ensure that no two items are assigned the same identifier, i.e. for each pair $x_a, x_b$ of distinct items produced by $p_j$, $a \ne b$. This can be easily achieved by concatenating the party's identifier $j$ with a local counter that is incremented for each new item.


\subsection{Event 2: Shipped Item} \label{s:event2}
When a party $p_s$ finishes the manufacturing processes of an item $x_i^{stage(p_s)}$ and supplies it to a buyer $p_b$ operating at the next stage, the procedure $shipItem(b, i)$ is invoked. Likewise $registerItem()$, this method simply tracks in the blockchain that the shipping of item $x_i$ from party $p_s$ to party $p_b$ has taken place. At line~\ref{l:ship} of module~\ref{m:tracking_system}, all the relevant shipping information are included in the key, i.e. in the first parameter of set operation, to make it easier retrieving shipping data. The value, i.e. the second parameter of set operation, is not significant and is set to $i$ by convention. Indeed, when querying the blockchain on whether a certain shipping of item $x_i$ from party $p_s$ to party $p_b$ took place, it suffices to check that the value stored for the key $\langle shipped, s, b, i \rangle$ is not null.

\subsection{Event 3: Delivered Item} \label{s:event3}
When an item is delivered to a party $p_b$ from a supplier $p_s$ operating at the previous stage, an integrity verification is carried out. 
Let $x_i^{stage(p_s)}$ be the delivered item. 
The party $p_b$ calls method $getChallenges(s,i)$ to obtain the $C$ challenges $crv$ to use and the expected responses from the CRD previously registered (event 1). 
Then, $p_b$ queries the PUF integrated with $x_i^{stage(p_s)}$ to obtain its actual responses $crv'$. Finally, $p_b$ calls the method $verifyItem(s, i, crv, crv')$ to compare $crv$ with $crv'$. At least $R$ out $C$ responses need to be correct for the validation to succeed. In this way the verification procedure can tolerate possible inconsistent responses returned by an intact PUF (see section~\ref{s:puf}). The verification outcome is finally tracked in the blockchain.

The $getChallenges(s,i)$ method verifies first the presence in the blockchain of the corresponding shipping transaction, and a $no\_ship$ alert is raised in negative case (line~\ref{l:no_ship}). 
%
The CRD is retrieved from the storage by using the item's identifier (line~\ref{l:get_crd}~\footnote{In this line, the operator $\pi_i$ is used to select the $i^{th}$ component of what the $get$ operation returns. Indeed, in the corresponding $set$ operation at line~\ref{l:set_crd}, the stored value has two components: the first one is the party's identifier and the second one is the CRD itself.}). If no corresponding CRD is found, a $no\_crd$ alert is raised (line~\ref{l:no_crd}). The challenge-response pairs for $p_b$ are retrieved and decrypted ($\overline{crd_i[p]}$ at line~\ref{l:select_challenge}). 
Finally, the chosen pair is tracked in line~\ref{l:declare_verification}.

The $verifyItem$ method checks whether the $C$ responses $r_k'$ obtained from the PUF queried with the challenges $c_k$ correspond to those expected, i.e. $r_k$.
Depending on whether at least $R$ out of $C$ responses match, the proper tracking information is stored in the key-value storage (lines~\ref{l:succeeded} and~\ref{l:failed}). 



\section{Security Analysis} \label{s:security_analysis}


In this section we discuss what a malicious party $p_A$ operating at stage $stage(p_A)$ can do and how our proposed tracking mechanism would react. We first define the relevant attacks an adversary may launch in section~\ref{s:attacks_definition}, based on the threat model introduced in section~\ref{s:threat_model} and the tracking system proposed in section~\ref{s:tracking}. Then, in section~\ref{s:attacks_analysis} we analyse the response of our tracking system to each of the identified attacks and whether it succeeds in coping with them.

\subsection{Attacks Definition} \label{s:attacks_definition}

According to the threat model introduced in subsection~\ref{s:threat_model}, the adversary $p_A$ can operate at different levels. As it cannot collude with any other party nor control their resources, attacks at \textit{supply chain level} are not relevant. At \textit{item level}, $p_A$ has several options. The basic one is to just forge an intact item before supplying it to another buyer (attack~\ref{a:forge_item}):

\begin{attack}
	\label{a:forge_item}
	The adversary $p_A$ tampers with an item received from an honest supplier and delivers it to an honest buyer at the next stage.
\end{attack}

%

If party $p_A$ works at stage $0$, the adversary can tamper with an item before its PUF is fed with the required number of challenges to compute the corresponding CRD. In this way, the CRD stored in the blockchain matches the forged item (attack~\ref{a:stage_0}):

\begin{attack}
	\label{a:stage_0}
	The adversary $p_A$ tampers with an item at stage $0$ before its CRD is generated and delivers it to an honest buyer at the next stage.
\end{attack}

The adversary can tamper with an item just after the delivery and before it gets verified by the tracking system. To avoid any attribution, $p_A$ can blame the corresponding supplier for the shipping of a counterfeit item (attack~\ref{a:blame_supplier}): 

\begin{attack}
	\label{a:blame_supplier}
	The adversary $p_A$ claims that an item it received from an honest supplier has been tampered with.
\end{attack}

At \textit{blockchain and smart contract execution environment} levels, the adversary can try to compromise the application logic of the smart contract or the data stored in the blockchain by properly instructing the local node $n_A$, i.e. node $n_A$ becomes byzantine (attack~\ref{a:byzantine}).

\begin{attack}
	\label{a:byzantine}
	The adversary $p_A$ alters the behaviour of the local node $n_A$, i.e. node $n_A$ becomes byzantine.
\end{attack}

The layer between supply chain business processes and tracking system is an additional attack surface to consider (see section~\ref{s:tracking}). At this level, the adversary can compromise the way smart contract methods are invoked, e.g. by using maliciously modified parameters or by not calling a method at all (attack~\ref{a:methods}): 

\begin{attack}
	\label{a:methods}
	The adversary $p_A$ alters how methods of the tracking system smart contract are called.
\end{attack}


\subsection{Attacks Analysis} \label{s:attacks_analysis}
for each of the five attacks identified in the previous subsection, we provide an analysis of how the proposed tracking system reacts.

\medskip \noindent \textbf{Analysis of Attack~\ref{a:forge_item}.} 
In this scenario, party $p_A$ tampers with an item $x_i^{stage(p_A)-1}$ received by an honest supplier $p_s$. Since the supplier is honest, we assume that $x_i^{stage(p_A)-1}$ has not been forged yet. 
The tampered item is supplied to another honest party $p_j$ at stage $stage(p_A)+1$. As $p_j$ is honest, it complies with the tracking mechanism described in section~\ref{s:tracking}, hence it retrieves its $C$ challenge-response pairs $crv$ from the $kvs$, queries the PUF $puf_i^{stage(p_A)}$ to compute the corresponding responses $crv'$ and calls the method $verifyItem(A, i, crv, crv')$ of the tracking system. 

We can assume that $p_A$ stored the correct tracking information regarding the shipping of $x_i$, otherwise an alert discrediting $p_A$ would be raised by $p_j$ (module~\ref{m:tracking_system}, line~\ref{l:no_ship}). We can also assume that the correct CRD of $x_i$ has been stored in the storage, indeed in this scenario $p_A$ does not operate at stage $0$, so the party which produced $x_i$ is honest. 
With reference to module~\ref{m:tracking_system}, this means that the checks at lines~\ref{l:check_ship} and~\ref{l:check_crd} are positive and the $C$ PUF responses in $crv'$ can be compared against those in $crvr$. From the properties expressed by equations~\ref{eq:tampering_check_positive} and~\ref{eq:tampering_check_negative}, and by the fact that $x_i^{stage(p_A)}$ has been tampered with, it follows that, with high probability, less than $R$ out $C$ responses match, hence an alert is raised (line~\ref{l:failed}) to notify the detection of a counterfeit item supplied by $p_A$. The accuracy of this forgery detection mechanism clearly depends on the choice of $R$. In section~\ref{s:evaluation} we show an experimental evaluation where $R$ is tuned to maximise the probability that counterfeits are recognised and minimise the chances that intact items are mistaken for forged.

It is to note that the challenge-response pairs that will be used for the verification are known by the intended party only, hence an adversary could not discover them in advance and build a model to be used to implement a clone.


\medskip \noindent \textbf{Analysis of Attack~\ref{a:stage_0}.}
If $p_A$ operates at stage $0$ and tampers with an item $x_i^0$, then there are two cases. If the counterfeiting occurs after the invocation of $registerItem()$ method, 
then this attack is equivalent to attack~\ref{a:forge_item} and the forgery is detected by the buyer of $x_i$ at stage $1$. Otherwise, if the tampering is made before and the stored CRD $crd_i$ accurately corresponds to $puf_i^0$, then this attack cannot be detected by the proposed tracking mechanism. 

\medskip \noindent \textbf{Analysis of Attack~\ref{a:blame_supplier}.} 
In this case, $p_A$ receives an authentic item $x_i^{stage(p_A)-1}$ from a supplier $p_j$, tampers with it and invokes $verifyItem(j, i, c, r, r')$ with $r \ne r'$. The reason why the adversary may want to launch this kind of attack is to discredit $p_j$. 
%
Likewise attack~\ref{a:forge_item}, the tracking system in the end succeeds in detecting the counterfeiting. Anyway, it fails in identifying the counterfeiter party, because the latter is identified by the first parameter of $verifyItem$, i.e. $j$. 

\medskip \noindent \textbf{Analysis of Attack~\ref{a:byzantine}.} 
The attacker can make the local blockchain node $n_A$ behave arbitrarily, i.e. it becomes a byzantine node, with the aim of 
compromising data stored in the blockchain or the application logic encoded in the smart contract of the tracking system.
By design, according to the model presented in section~\ref{s:blockchain_model}, in a blockchain with $N$ nodes the adversary should control at least $\lfloor \frac{N-1}{3} \rfloor + 1$ nodes to compromise the consensus, hence if there are at least $4$ parties in the supply chain, each with its own local blockchain node, then this attack cannot succeed.

\medskip \noindent \textbf{Analysis of Attack~\ref{a:methods}.} 
The adversary can interact with the methods provided by the tracking system differently from what expected. In the specific, $p_A$ can either invoke a method when it should not, or avoid to call a method at all, or purposely specify wrong values for methods parameters.
As explained in section~\ref{s:tracking}, an adversary cannot call any method on behalf of another party, hence $p_A$ can only operate on the methods it is expected to invoke. 

If $p_A$ operates st stage $0$, it can intentionally avoid to store the CRD for item $x_i$, i.e. it can skip calling $registerItem()$ method. The motivation could be to prevent forgery checks from taking place and indeed such a goal can be partially achieved by the attacker. Anyway, the honest party $p_j$ receiving $x_i$ from $p_A$ easily discovers that the required CRD $crd_i$ is missing (line~\ref{l:check_crd}) and raises an alert (at line~\ref{l:no_crd}, with parameters $no\_crd, A, j, i$). Although no forgery can be actually detected in this way, that alert marks $x_i$ as a suspicious item and $p_A$ as a disreputable party because it did not store the CRD. 

If $p_A$ does not call method $shipItem()$ when expected, then the next party receiving the corresponding item $x_i$ detects this anomaly at line~\ref{l:check_ship} and consequently raises an alert at line~\ref{l:no_ship}, which again explicitly points at $p_A$ as the party responsible for this misbehaviour.

Avoiding the execution of $getChallenge()$ and $verifyItem()$ methods would bring no advantage to the adversary, with respect to its goal (see section~\ref{s:threat_model}) of introducing counterfeited products without being detected.

Altering the parameters used for either $registerItem()$ or $shipItem()$ method has the same effect of not calling them at all. If instead parameters are altered for $getChallenge()$ or $verifyItem()$ method, the the integrity check fails, which leads to the same consequences of attack~\ref{a:blame_supplier}.

\section{Experimental Evaluation} \label{s:evaluation}

We implemented a prototype of the proposed solution to verify the technical feasibility of the integration of blockchain and PUF, and to assess the reliability of PUF technology to accurately detect counterfeit.
We used HyperLedger Fabric~\footnote{Hyperledger Fabric (\url{https://www.hyperledger.org/projects/fabric})} to implement the consortium blockchain and the smart contract execution environment (see sections~\ref{s:blockchain_model} and~\ref{s:smart_contract_model}). We chose this platform because it is one of the most stable and well documented platforms for consortium blockchains. The tracking system $\mathcal{TS}$ defined in module~\ref{m:tracking_system} has been coded as a Fabric chaincode. A 4 bit sequential ring oscillator architecture~\cite{wachsmann2014physically} PUF has been synthesised and implemented on 17 separate Zynq Zybo 7000 FPGA boards~\cite{crockett2014zynq}.

The interface between the tracking system and the PUFs has been implemented as a Java application. The communication with PUF has been done using RXTXComm~\footnote{RXTXComm (\url{https://seiscode.iris.washington.edu/projects/rxtxcomm})}, a library which makes use of Java Native Interface (JNI~\footnote{JNI (\url{https://docs.oracle.com/javase/8/docs/technotes/guides/jni/})}) to provide a fast and reliable method of communication over serial ports. The communication at PUF side has been encapsulated in a dedicated module which used General Purpose Input Output (GPIO) as Tx and Rx pins for Universal Asynchronous Receiver/Transmitter (UART) serial communication.

We first describe how we tuned the PUF (section~\ref{s:puf_tuning}), then we describe the use case we tested and what results we obtained (section~\ref{s:prototype_test}).

\subsection{PUF Tuning} \label{s:puf_tuning}

The tuning of PUFs consisted in choosing the right value of parameter $R$, i.e. how many responses out of $C$ need to be correct for the validation to succeed, where $C$ is the number of unique challenges sent to the PUF. We set $C$ to 10.

We first generated the CRD for all the 17 PUFs by collecting a large number of challenge-response pairs for each PUF (more than 21000 pairs). We then randomly selected 3 out of the available 17 PUFs for tuning, while the others were used for the prototype test (section~\ref{s:prototype_test}). We refer to those 3 PUFs as the tuning PUFs. Challenges drawn from CRD data of all the PUF have been sent to the tuning PUFs to collect the correspondent responses. The resulting dataset has been used to find a value of $R$ that guarantees that each tuning PUF (i) passes the validation when stimulated with its own CRD and (ii) fails the validation when stimulated with CRD of any of the other 16 PUF.

Each tuning PUF has been stimulated with $C=10$ unique challenges from each of the 17 PUFs (hence including itself) for 15 times. For each batch of $C$ challenge-response pairs, different values of $R$ has been tested, ranging from 5 to 9, and the corresponding validation outcome has been recorded. The metrics of interest for the tuning are
\begin{itemize}
	\item True Admission Rate (TAR): rate of successful validations when the tuning PUF is validated against its own CRD;
	\item False Admission Rate (FAR): rate of successful validations when the tuning PUF is validated against the CRD of another PUF;
	\item True Rejection Rate (TRR): rate of failed validations when the tuning PUF is validated against the CRD of another PUF;
	\item False Rejection Rate (FRR): rate of failed validations when the tuning PUF is validated against its own CRD;
\end{itemize}

The ideal situation is when TAR and TRR are 1 while FAR and FRR are 0.


\begin{figure*}
	\centering
	\subfigure[Tuning PUF 1]{\label{f:tuning_puf_1}\includegraphics[width=4.45cm]{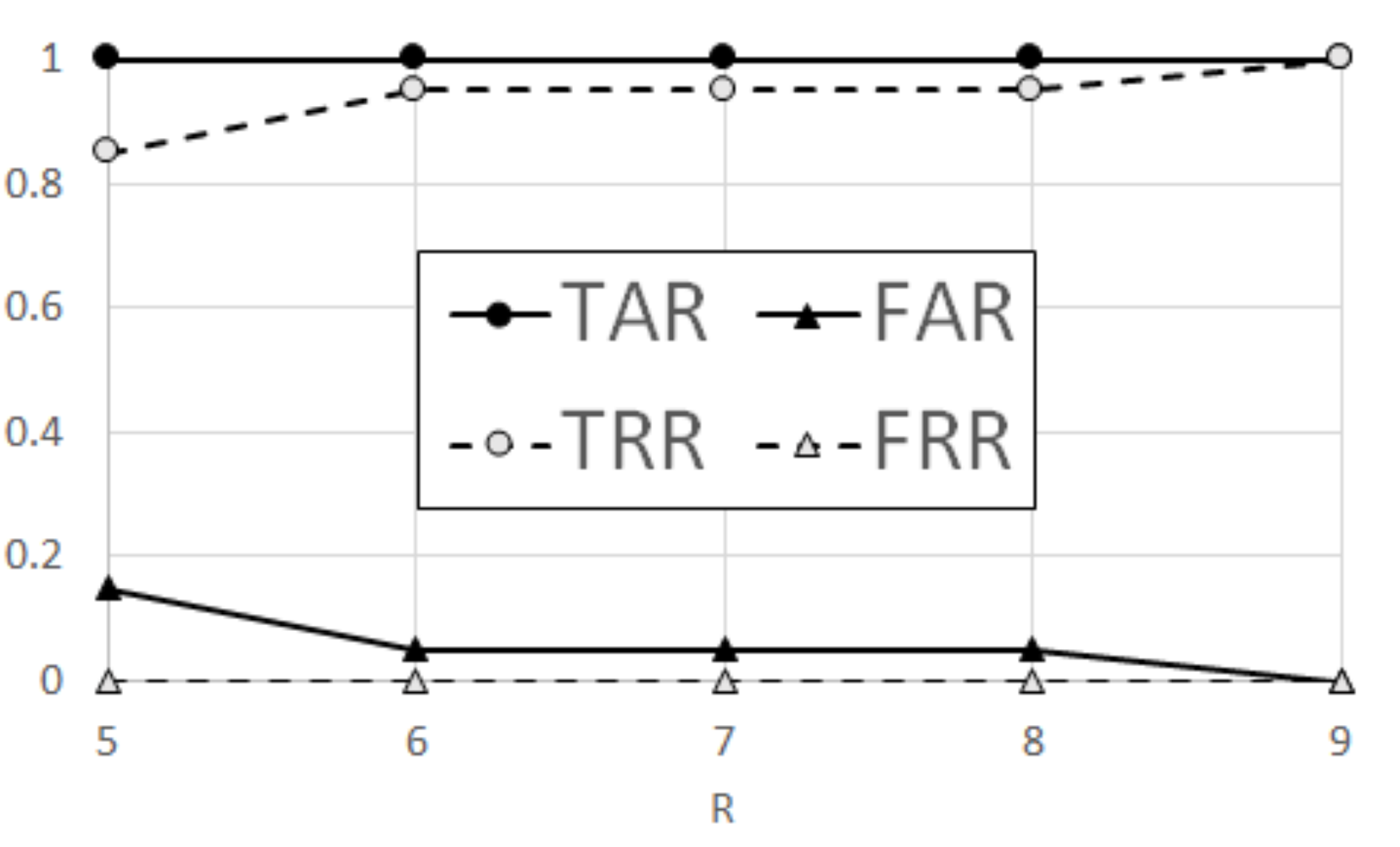}}
	\subfigure[Tuning PUF 2]{\label{f:tuning_puf_2}\includegraphics[width=4.45cm]{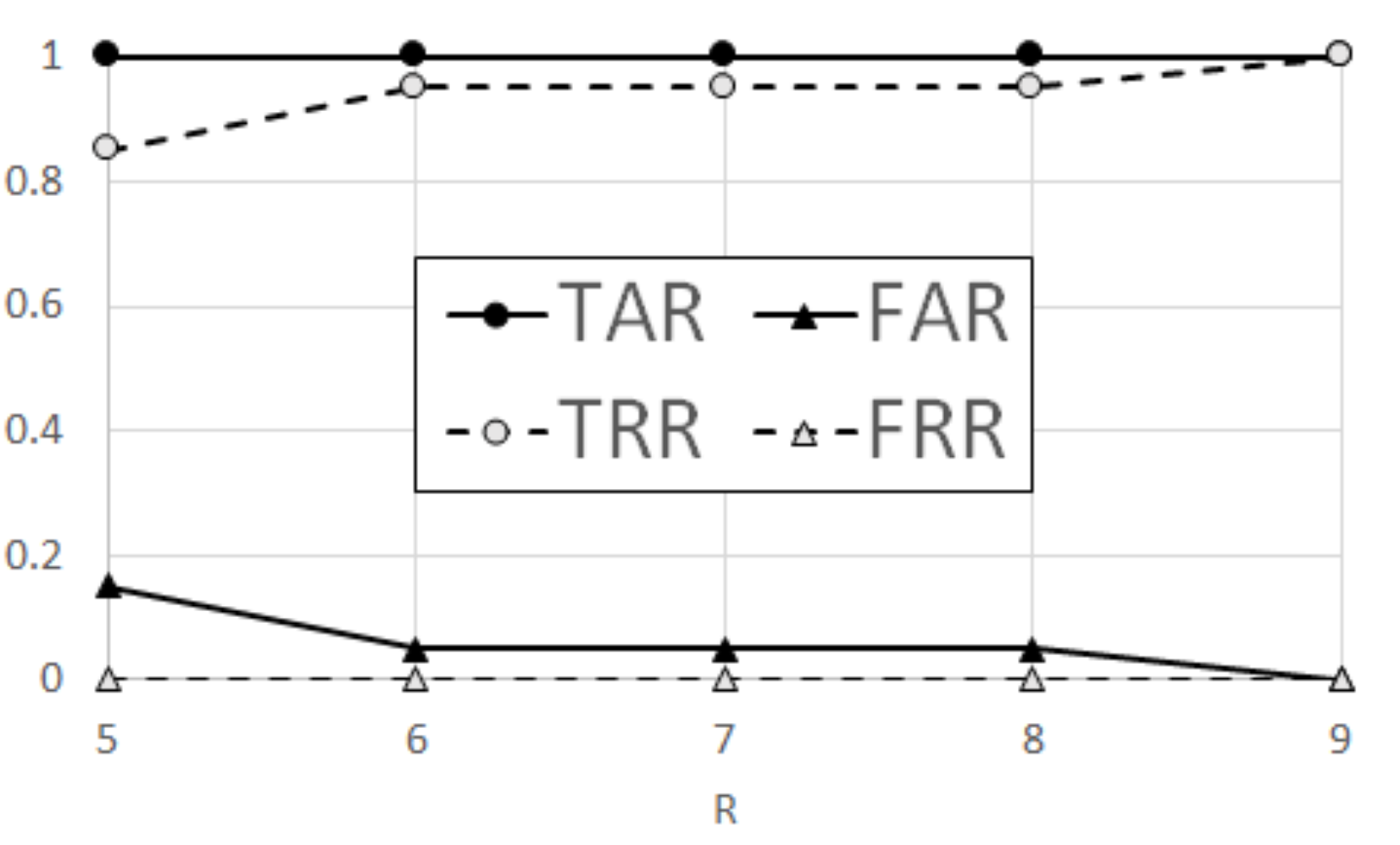}}
	\subfigure[Tuning PUF 3]{\label{f:tuning_puf_3}\includegraphics[width=4.45cm]{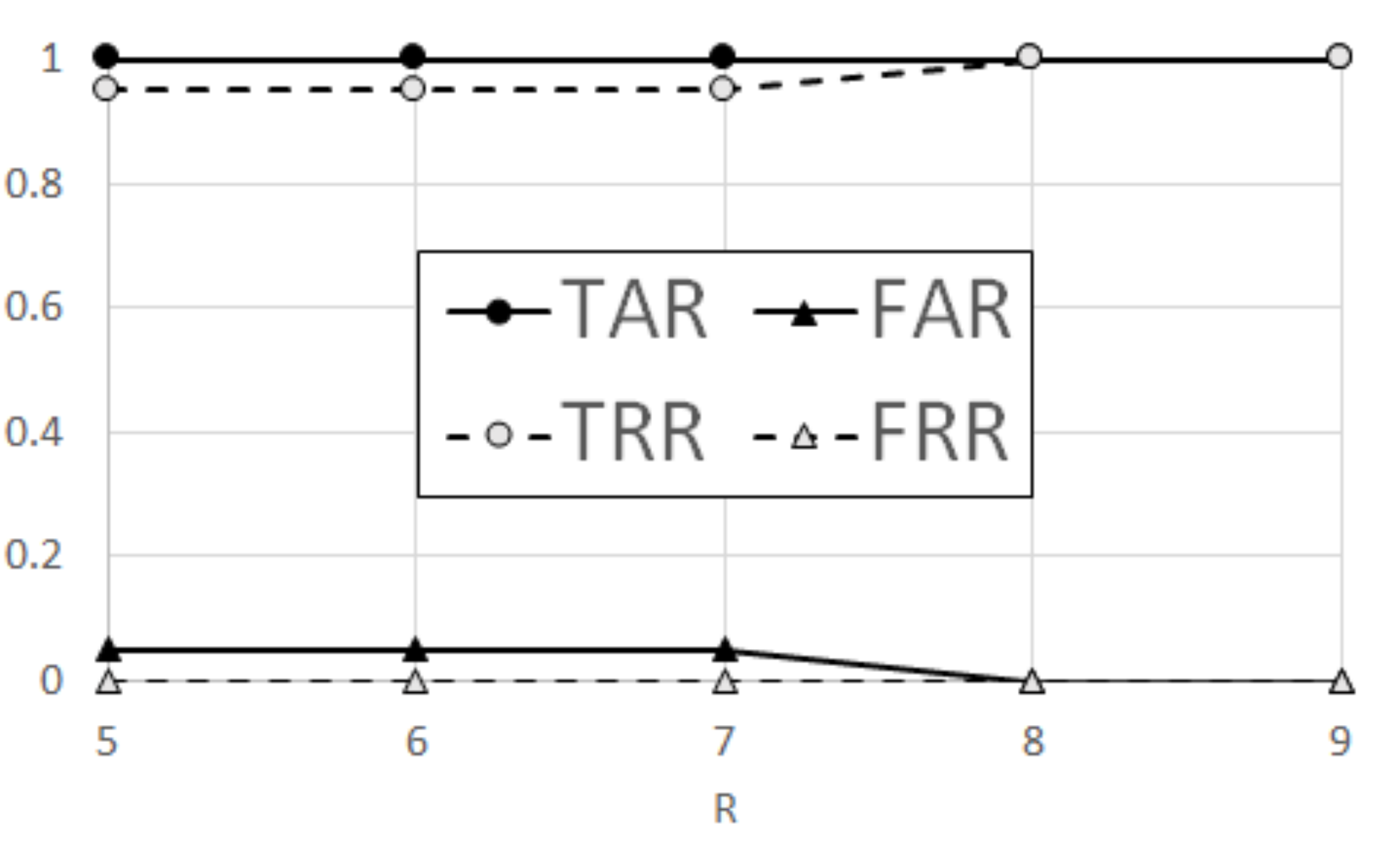}}
	\caption{Tuning of 3 PUFs, R is varied between 5 and 9 out of 10 and the corresponding value of metrics TAR, FAR, TRR, FRR are shown.}
	\label{f:tuning_pufs}
\end{figure*}

Figure~\ref{f:tuning_pufs} shows the values of those metrics for $R$ varying from 5 to 9 (out of 10) for the three tuning PUFs. It can be noted that TAR is always 1 and FRR always 0, which means that the tuning PUFs are successfully validated all the times their own CRD is used. When the validation is based instead on CRD of a different PUF, sometimes tuning PUFs still pass the validation. This happens because the functions computed by different PUFs can overlap for certain challenges. Figure~\ref{f:tuning_pufs} shows that the probability that this occurs (i.e. FAR) decreases as R grows, and that with $R=9$ FAR is 0 (and TRR is 1) for all the 3 tuning PUFs. Hence, for the prototype test, the validation of a PUF is considered successful if at least 9 out of 10 responses match those stored in the corresponding CRD.

\subsection{Prototype Test} \label{s:prototype_test}

\begin{figure}[h]
	\includegraphics[width=8cm]{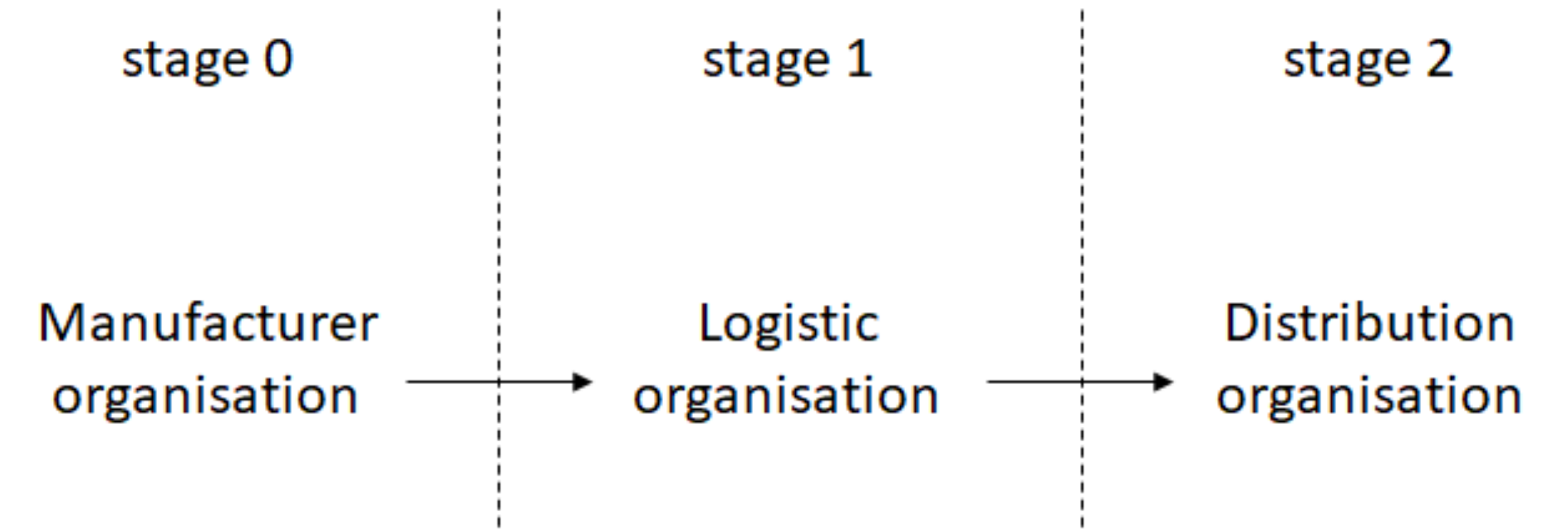}
	\centering
	\caption{Supplier-buyer relationships in the prototype.}
	\label{f:prototype}
\end{figure}

We developed a prototype with three organisations: manufacturer, logistic and distribution. The corresponding supplier-buyer relationships are depicted in figure~\ref{f:prototype}. We considered two cases: when no adversary is present and when the logistic organisation is malicious and tampers with the items supplied by the manufacturer before delivering them to the distribution organisation.

We used the other 14 PUFs for the prototype test, 8 for the case where no party is malicious and 6 for the case where the logistic organisation is the adversary. In the latter case, the manufacturer delivers 3 PUFs to the logistic organisation, which replaces each of them using the other 3 PUFs and deliver them to the distribution organisation.

When there is no adversary, all the 8 PUFs pass the validation both at the logistic and at the distribution organisation, hence the TAR is 1 and FRR is 0. When instead the logistic organisation replaces the the three PUFs, all of them fail the validation at the distribution organisation, therefore the FAR is 0 and TRR is 1.

These preliminary results are promising to prove both the technical feasibility and the effectiveness in counterfeit mitigation of the proposed tracking system.

\section{Discussion} \label{s:discussion}
This section discusses several key aspects of the proposed solution, pointing out key limitations and main research directions to investigate as future work: the results of the security analysis (section~\ref{s:discuss_security_analysis}), the issues of implementing a PKI infrastructure for a consortium blockchain (section~\ref{s:discuss_pki}), the limitations of the chosen threat model (section~\ref{s:discuss_threat_model}), the feasibility of embedding PUFs within the items to track (section~\ref{s:discuss_puf_embedding}), possible privacy issues when sharing data among parties through the blockchain (section~\ref{s:discuss_privacy}), observations on consortium blockchain performance and scalability (see section~\ref{s:discuss_performance_and_scalability}) and, finally, considerations on the costs associated with adopting the proposed solution in real supply chains (section~\ref{s:discuss_platform_costs}).

\subsection{Security Analysis Results} \label{s:discuss_security_analysis}
The results of security analysis presented in section~\ref{s:security_analysis} show the capability of the proposed tracking system to be effective against the identified attacks. Any attempt to counterfeit items (attack~\ref{a:forge_item}) is correctly detected and attributed to the right malicious party.

If the adversary operates at stage $0$ and tampers with the item before the corresponding CRD is built and stored in the blockchain (attack~\ref{a:stage_0}), then the tracking system fails to detect the forgery. This derives trivially from relying on the CRD itself to be the trust root of the whole counterfeit detection mechanism. Enhancing the proposed approach to cover threats happening before CRD generation is one of our main future work.

In case the attacker forges a received item just before the integrity checks (attack~\ref{a:blame_supplier}), the tracking system succeeds in spotting the counterfeiting but fails in the attribution, i.e. the honest supplier of the adversary is held accountable. Improving the accuracy in identifying the malicious party is another relevant future extension for this work.

The other attacks at software level, to make a blockchain node byzantine (attack~\ref{a:byzantine}), or at the interface between supply chain business processes and tracking system (attack~\ref{a:methods}), have been shown to be not effective. On the one hand, this derives from by-design security properties provided by blockchain-based systems, indeed using PBFT-like consensus algorithms allows to tolerate a single byzantine node when the blockchain includes at least four nodes (attack~\ref{a:byzantine}). On the other hand, the tracking system prevents an adversary from invoking smart contract methods on behalf of a different party, so attacks based on altering how methods are called (attack~\ref{a:methods}) are not relevant.

\subsection{PKI Infrastructure for Consortium Blockchains} \label{s:discuss_pki}
The proposed tracking system relies on a consortium blockchain (see section~\ref{s:blockchain_model}), which in turn requires a reliable PKI to obtain the relationships between parties' identities and public keys. These certificates are issued when the platform is setup at the beginning and when the supply chain membership changes. From a security perspective, the PKI is a single-point-of-failure, i.e. an adversary may target the PKI to take over the whole blockchain, and thus the tracking system.


This problem has been already addressed in literature. For example, there exist proposed solutions based on blockchain to decentralise the PKI so as to make it much more resistant to cyber attacks~\cite{Al-Bassam:2017:SSC:3055518.3055530, fromknecht2014decentralized}, and provide attack tolerance guarantees comparable to those already provided by the tracking system. These solutions are based on public blockchains, which may introduce privacy issues. Other approaches have been proposed for privacy-preserving blockchain-based PKI, such as PB-PKI~\cite{axon2016pb}. The integration of the tracking system with this type of PKI is out of the scope of this paper and is left as future work.

\subsection{Threat Model Limitations} \label{s:discuss_threat_model}
The list of attacks identified in section~\ref{s:attacks_definition} depends tightly on the threat model introduced in section~\ref{s:threat_model}, which in turn derives from two main assumptions: (i) there is a single adversary and (ii) it controls exactly one party. It can be reasonable to consider the implications of relaxing those assumptions and identify what additional attack scenarios may arise when an adversary can control more parties or when more adversaries are active, either independently or by colluding among themselves.

We can expect that a security analysis of the proposed tracking system against such a stronger attack model would point out further vulnerabilities. However, this analysis should be integrated with a risk assessment to measure the likelihood of more advanced attacks, and should estimate out to what extent they can be considered reasonable. Taking into account wider threat models is an additional potential future work.

\subsection{Embedding PUFs within Items to Track} \label{s:discuss_puf_embedding}
The effectiveness of tracking items by using PUFs strictly depends on how easily an adversary can forge items without affecting the PUFs themselves. If a PUF can be removed from an item and embedded within a different one, then the whole counterfeit detection mechanism is flawed. In the end, this boils down to preliminarily check whether it is technically feasible to embed PUFs within items in such a way that all the properties of the PUF-equipped item model hold true (see section~\ref{s:puf_model}).

Electronic components are items where PUFs can be be easily and cheaply implanted by integrating PUF circuitry inside the component circuitry, ensuring that PUFs cannot be removed and replaced. Hence, the approach we propose fits well with integrated circuits and IoT devices supply chains.

\subsection{Privacy Issues} \label{s:discuss_privacy}
Although the network of companies involved in the supply chain should be made as transparent as possible to enhance visibility, organisations can be legitimately reluctant to disclose their own supplier network and procurement history to other, possibly competitor firms. What information should be shared needs to be adjusted according to this kind of confidentiality requirements, on a case by case basis. An important applied research direction to investigate, for each target supply chain market, concerns this trade-off between privacy and scope, with the aim to find the sweet spot where information on supplier network and procurement history can be shared smoothly.

A general approach to address those privacy issues is to make each transaction only visible to a specific subset of parties. With reference to our prototype implementation, Hyperledger Fabric provides the concept of \textit{channels} to establish between subsets of nodes. 
A transaction can be associated to a specific channel to ensure only the nodes in that channel can see its content. Our prototype can be enhanced with privacy-preserving techniques by relying on Fabric channels.

\subsection{Performance and Scalability} \label{s:discuss_performance_and_scalability}
While public permissionless blockchains like Ethereum's are known to provide limited performances in terms of transaction latency and throughput, consortium blockchain can commit thousands of transactions per seconds with subsecond latency~\cite{bessani2014state}, also in WAN settings~\cite{sousa2015separating}. 
In terms of scalability, BFT-tolerant algorithms have been proposed in literature that can scale to tens of nodes with minor performance penalties~\cite{vukolic2015quest, junqueira2011zab}, which matches realistic supply chain setting including tens of different organisations.

\subsection{Platform Integration Costs} \label{s:discuss_platform_costs}
Each supply chain works according to specific business processes which may differ significantly from market to market. On the one hand, pinpointing the right abstraction level for the interface provided by the tracking system is crucial to increase the cases where it can be integrated. On the other hand, the integration with those business processes deserves a deeper analysis in terms of security, to figure out whether additional cyber threats can be identified at those integration points (see attack~\ref{a:methods} in section~\ref{s:attacks_definition}), and cost-effectiveness, to quantify whether and to what extent the benefits of counterfeiting mitigation outweigh the costs to accomplish such a large-scale integration.

In terms of cost-effectiveness, it is to note that relying on consortium blockchains rather public permissionless blockchains allows to cut any cost due to the fees to pay when submitting transactions. Indeed, while supply chain tracking solutions based on Ethereum have a per-transaction cost (e.g. see Negka et al.~\cite{negka2019employing}), submitting transactions in Hyperledger Fabric is totally free.

\section{Conclusion} \label{s:conclusion}

In this paper we design a tracking system to mitigate counterfeits in supply chains of physical products. The solution we propose is based on blockchain and smart contract technologies to provide high availability and strong tolerance against integrity attacks against stored data and application logic. We rely on physically unclonable functions to uniquely identify and accurately track items along the supply chain. We validate our solution against a specific threat model and find out that it is effective to counter the identified attacks, but (i) an adversary operating at the first stage of the supply chain can bypass the anti-counterfeit mechanism and (ii) an adversary receiving an item from a honest supplier can tamper with that item and blame the supplier itself for the forgery. Finally, we implemented and tested a prototype of the proposed tracking system to prove it is technically feasible and accurate in correctly validating both intact and forged items.

In addition to investigate possible solutions to the limitations discovered in the security analysis, other future work include the integration of a reliable PKI infrastructure within the tracking system and the implications of  considering a stronger threat model.

\bibliographystyle{elsarticle-num}
\bibliography{references}

\end{document}